\documentclass[prb,twocolumn,showpacs,preprintnumbers]{revtex4}

\usepackage{graphicx}

\begin{document}

\title
{ The Coulomb bridge function and the Pair-distribution functions
 of the 2-dimensional
 electron liquid in the quantum regime.
}

\author
{
 M.W.C. Dharma-wardana}
\affiliation{
National Research Council of Canada, Ottawa, Ontario, Canada, K1A 0R6
}
\email[Email address:\ ]{chandre.dharma-wardana@nrc-cnrc.gc.ca}

%
\date{\today}

\begin{abstract}
The electron-electron pair distribution functions (PDF) of the
2-D electron fluid (2DEF) in the quantum regime (at $T$=0)
 are calculated using
a classical-map-hyper-netted-chain (CHNC) scheme and compared with
currently available Quantum Monte-Carlo (QMC) simulations in the
coupling range $r_s$=1 to 50. We iteratively extract the
bridge function of the ``equivalent'' classical 2-D liquid
in the quantum regime. These bridge
functions $B(r)$ are relatively insensitive to spin-polarization
effects. The structure of the bridge functions changes
significantly for $r_s>6$, suggesting the onset
of strongly correlated clusters.
The new $B(r)$, appropriate for the
long-range Coulomb potential, can be used
to replace the hard-sphere $B(r)$ previously used in these
calculations. They provide accurate classical
 representations of the
QMC-PDFs even at very strong coupling, and probably
at finite-$T$ near $T=0$.
\end{abstract}
%
\maketitle
\section{Introduction}
The pair-distribution functions (PDFs) of strongly-coupled electron fluids
contain all the physical information associated with the ground-state
 static properties of such systems. Exchange-correlation
energies, phase-transitions, and Fermi-liquid parameters like
 the effective mass $m^*$, and the spin-susceptibility enhancement
 ($g^*$) can all be evaluated from the PDFs, as discussed below.
 The static local-field
corrections to the response functions can also be addressed via these
PDFs.

The PDFs are usually determined by quantum Monte-Carlo (QMC) simulations,
since standard many-body methods become unreliable for densities 
where the electron-sphere radius $r_s$ exceeds unity. The $r_s$ parameter
is also the ratio of the mean Coulomb energy and the Fermi energy, and hence is a measure of the coupling strength.
QMC simulations for the 2D electron system were first published by
Tanatar and Ceperley~\cite{tancep}, and more recently by
 Attaccalite et al.~\cite{atta}, and by 
Drummond and Needs~\cite{Drum2009}. A transition from the
paramagnetic state to the ferromagnetic phase was predicted to
occur at $r_s\sim$26 by Attaccalite et al., while weak-coupling
theories (Hartree-Fock, RPA) predicted such transitions
already at very low values of $r_s$. In contrast, Tanatar and Ceperley,
 as well as the most recent QMC work by Drummond et al., find no
such phase transition, where the ferromagnetic state is very close
in energy, but the paramagnetic state remains the ground state.
Indeed, direct comparisons of the QMC-PDFs of Attaccalite et al.,
with the more accurate PDFs of Drummond et al., show  slight
differences which show the need to be cautious about undue claims
of final accuracy.

QMC methods have also been used to calculate Fermi-liquid 
parameters~\cite{atta} like $m^*$ and $g^*$, but these, and especially the $m^*$ calculation~\cite{drum2}
may require further effort before a consensus is reached.

Besides QMC, one other method~\cite{prl2}
 available for the calculation of PDFs of
quantum (e.g., electron) fluids at arbitrary coupling, temperature and
spin polarization is based on considering a classical charged fluid
at an assigned classical-fluid temperature $T_{cf}$, selected
to reproduce the
correct correlation energy of the quantum fluid which has two spin
species $\alpha, \beta$.
The non-interacting
PDFs $g_{\alpha,\beta}^0(r)$ of the classical charged fluid are  
formulated to agree with 
the analytically known non-interacting  $g_{\alpha,\beta}^0(r)$
of the quantum fluid. This is done by introducing an effective
potential known as the Pauli exclusion potential $P_{\alpha,\beta}$
to exactly reproduce the Fermi hole in the parallel-spin
$g^0(r)$. Once an ``equivalent'' classical map of the quantum
fluid is constructed, its PDFs are obtained using a classical
integral equation, namely, the hyper-netted-chain (HNC)
 equation modified to include 
bridge corrections. This classical-map hyper-netted chain procedure was
called the CHNC, and we showed that the method was surprisingly
accurate. The classical-fluid temperature of the quantum fluid at $T=0$ was called the ``quantum temperature'' $T_q$. 
In fact, using {\it only} the correlation energies tabulated by
Tanatar and Ceperley as the inputs, we constructed the effective
quantum temperature $T_{q}$ as a function of $r_s$,
 and showed that
accurate PDFs of the quantum fluid could be
calculated at arbitrary $r_s$ via the classical map~\cite{prl2}.
Balutay and Tanatar~\cite{balutay} also presented a closely similar
temperature map.
The advantage of the CHNC approach is that it is numerically very simple,
and makes many properties (spin-dependent properties, Fermi-liquid parameters,
local-field corrections, finite-$T$ results, etc.) easily computable
with negligible effort. Thus the product $m^*g^*$ can be
evaluated~\cite{gstrmstr} from the second derivative of the correlation energy
with respect to the the spin-polarization $\zeta$.
The effective mass $m^*$ can be obtained~\cite{gstrmstr}
from the second density-derivative of the free energy which
 can be evaluated from the temperature-dependent PDFs. 

However, the classical map of the 2D- electron fluid leads
 to a Coulomb fluid
for which the simple HNC is inadequate. The irreducible three-body
and higher-order contributions which are lumped into
 the bridge term $B_{\alpha,\beta}(r)$
are very difficult to calculate directly. 
The need for bridge functions appears in many
areas in the theory of Coulomb systems~\cite{Wrighton}.
Hence, as is customary~\cite{rosen}, in our earlier work
 we used
a hard-disk model, based on solutions of the Percus-Yevik
equation~\cite{yr2d}.  However, the availability of
an extensive set of QMC-generated PDFs for the 2D-electron fluid
presents the possibility of
extracting accurate Coulomb-adapted bridge functions $B(r)$.
These 
are implicit functions of the PDFs themselves~\cite{Poll-ash}.
Thus, in this paper we present an iterative procedure
for extracting the 2D-bridge functions appropriate to a
classical Coulomb fluid at a given $r_s$, and at the effective
classical-fluid  temperature
$T_{cf}$, and yielding the {\it quantum}
 PDF at the given $r_s$ and at $T=0$.


\section{HNC and CHNC methods}
\label{sec-hnc}
The HNC equation~\cite{hncref} and
its  generalizations, coupled with the  Ornstein-Zernike  equation
 have lead
to very accurate results for classical charged-particle interactions.
In the following we use indices $i,j$ which could be
spin indices or other species-identifying indices.
 The exact equations for the PDFs are of the form:
\begin{eqnarray}
g_{ij}(r)&=&e^{-\beta_{cf} \phi_{ij}(r)+N_{ij}(r)
+B_{ij}(r)} \\
N_{ij}&=& h_{ij}(r)-c_{ij}(r)\\
h_{ij}(r)&=&g_{ij}(r)-1 
\label{hnc}
\end{eqnarray}
Here $\phi_{ij}(r)$ is the pair potential between the
species $i,j$, and $N_{ij}(r)$ is the nodal function, while
$c_{ij}(r)$ is the direct correlation function connected to
$h_{ij}$ by the
Ornstein-Zernike (O-Z) equation. The pair potential $\phi_{ij}(r)$
is the the sum of a diffraction corrected Coulomb potential and the Pauli exclusion potential~(see Eq. 1-2 , ref~\cite{prl2}).

If the bridge function $B_{ij}(r)$ were set to zero we have
 the HNC approximation, adequate for systems where the kinetic
 energy dominates
strongly over the potential energy. The bridge function brings in
many-body cluster interactions beyond the diagrams of the
hyper-netted chain expansion.  A systematic investigation of the
 3-D one-component plasma
was given many years ago by Lado et al~\cite{rosen}. As the bridge
interactions involve many-internal interactions (within the cluster)
averaged over, they are somewhat insensitive to the exact
 form of the pair-interactions, and their spin-states.
 Thus it was shown~\cite{yr2d} that
the analytically available Percus-Yevik hard-sphere bridge function
could be used for most fluids, to closely reproduce the Monte-Carlo
PDFs available at the time. The hard-sphere radius of the
model bridge function was
in effect an optimization parameter in such
 approaches to liquid structure. 

The PDFs of the classical 2D electron fluid became relevant to 2D quantum
fluids when Laughlin~\cite{lau83} introduces his plasma mapping of
 2D fractional-quantum Hall (FQH) fluids to classical plasmas.
 The use of a suitable bridge function was found to be essential
 if high accuracy was to be obtained via the 
plasma map~\cite{fqhe-mac} for the hierarchy of FQH states.
 The  temperature of the classical fluid in Laughlin's
classical map is directly related to the filling factor, via the
 form of the many-body wavefunction proposed by Laughlin
(Laughlin used a simple HNC without bridge corrections). 
There is no external magnetic field in the 2D systems studied here.
It turns out that the temperature of the classical
 charged fluid whose PDF
 agrees with the electron (quantum) fluid at $T=0$
can be expressed~\cite{prl2} as a function of
the density parameter $r_s$. 

\begin{figure}
\includegraphics[width=8.8 cm, height=7.0 cm]{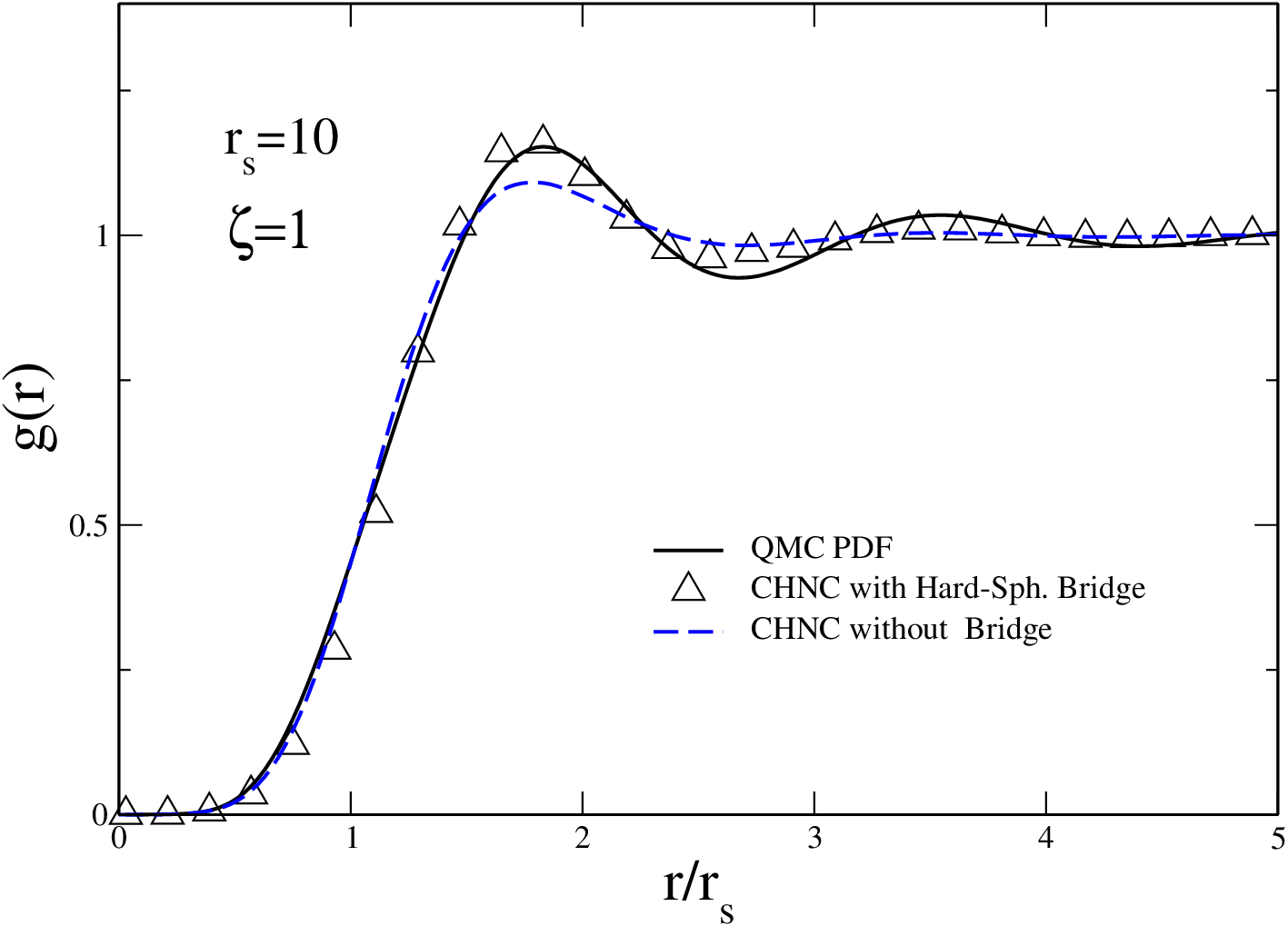}
 \caption
{(Online color) The QMC pair-distribution function of a fully spin polarized ($\zeta=1$)
electron fluid at $r_s=10, T=0$ is compared with those calculated from CHNC using
a hard-sphere bridge function and with no bridge function what so ever}
\label{comparison-rs10}
\end{figure}

We found~\cite{prl2} that the use of hard-sphere model bridge functions
gave good agreement around the first peak of the PDFs, while the more distant oscillations were weaker than those in the QMC-PDFs. The PDF of the
fully spin-polarized
$T=0$ electron fluid at $r_s=10$, calculated using CHNC without bridge,
and CHNC with a hard-sphere bridge function~\cite{prl2}, and the bench-mark QMC-PDF are
shown in Fig.~\ref{comparison-rs10}. 

The bridge function $B(r)$ is a
function of the PDFs themselves, and hence their extraction from
QMC simulation results is by no means obvious. However, given a choice for the effective pair potential and classical-fluid temperature
implicit in the CHNC $\beta\phi(r)_{ij}$, and the assumed
applicability of the modified-HNC equation and the O-Z equation,
a bridge function can be evaluated from the QMC-$g(r)$.
This evaluation requires separating out the long-range tails of the
pair-potentials, direct correlations etc., in $r$- and $k$- space,
using numerical Fourier transforms for the short range parts, analytical formulae for the long ranges parts and reassembling the results in $r$- and $k$-space. Since these mathematical machinery are already available in the algorithms for the CHNC, we present here a
simple iterative scheme equivalent to the above, but using the 
CHNC code itself.

The $B(r)$ functions obtained from such a procedure,
 together with the
$\beta\phi(r)$ of CHNC present a complete, accurate, classical
representation of the quantum PDFs obtained by QMC.
It is hoped that such bridge functions and classical maps can
then be used in
regimes where QMC simulations are not possible or easily available,
as discussed below.
\section{Extraction of the bridge function from QMC data}
\label{sec-bridge}
Given a density $n=1/(\pi r_s^2)$ specified by an $r_s$ value,
 the target QMC-PDFs that we use are those of
Attaccalite et al., as parametrized by Gori-Giorgi
et al.~\cite{Gori-giorgi}. We drop the species indices ${i,j}$
except when required, and indicate the PDFs from QMC and CHNC
by $g(r)$ and $g_{chnc}(r)$. 
Given the hypothesis that the QMC-PDFs can be represented
 by classical forms,
a nodal function and bridge function corresponding to the
 given target $g(r)$
should exist.  Thus
the QMC and the CHNC PDFs satisfy
\begin{equation}
\label{qmc-bridge}
log\left[g(r)\right]=-\beta\phi(r) + N(r)+B(r)
\end{equation}
Both $N(r)$ and $B(r)$ are
implicit functions of $g(r)$. 
 However, all the terms in the CHNC form
\begin{equation}
\label{chnc-bridge}
log\left[g_{chnc}(r)\right]=-\beta\phi(r) + N_{chnc}(r)+B_{chnc}(r)
\end{equation} 
are known. Also, it has been found from previous comparisons~\cite{prl2}
 of
$g(r)$ and $g_{chnc}(r)$ that they agree closely, even when $B_{chnc}(r)$
was taken from a hard-disk model. Hence we assume that $N(r)$ can be
replaced by $N_{chnc}$ as a first approximation.
Then we easily obtain an initial estimate of the bridge
function contained in the QMC-PDF.
\begin{equation} 
\label{iter-bridge}
B(r)=log\left[g(r)\right]-log\left[g_{chnc}(r)\right] + B_{chnc}(r)
\end{equation}
Thus, starting from the hard-sphere model of $B_{chnc}(r)$ we obtain a a standard mixing procedure to construct a new $ B_{chnc}(r)$,
and hence a new $g_{chnc}$, and so on. In the small-$r$ region the
value of $g(r)$ becomes negligibly small and hence the extraction of the
difference between two logarithms becomes numerically unsatisfactory. However, as
may be surmised from
Eq.~\ref{iter-bridge}, the calculated $g(r)$ are found to be
 insensitive 
to the form of $B(r)$ for $r < r_s$. Thus even a simple polynomial extrapolation
(connecting the $r/r_s < 1$ region with, say the region $1<r/r_s<1.5$)
 or
the hard-sphere model itself may be used. The iterative procedure
is even insensitive to slight discontinuities at the
connection point (although of course, discontinuities should
be avoided). These bridge functions will be called
 ``Coulomb bridge functions'', to distinguish them
from the hard-sphere bridge functions $B_{HS}(r)$.

\begin{figure}
\includegraphics*[width=8.5 cm, height=6.5 cm]{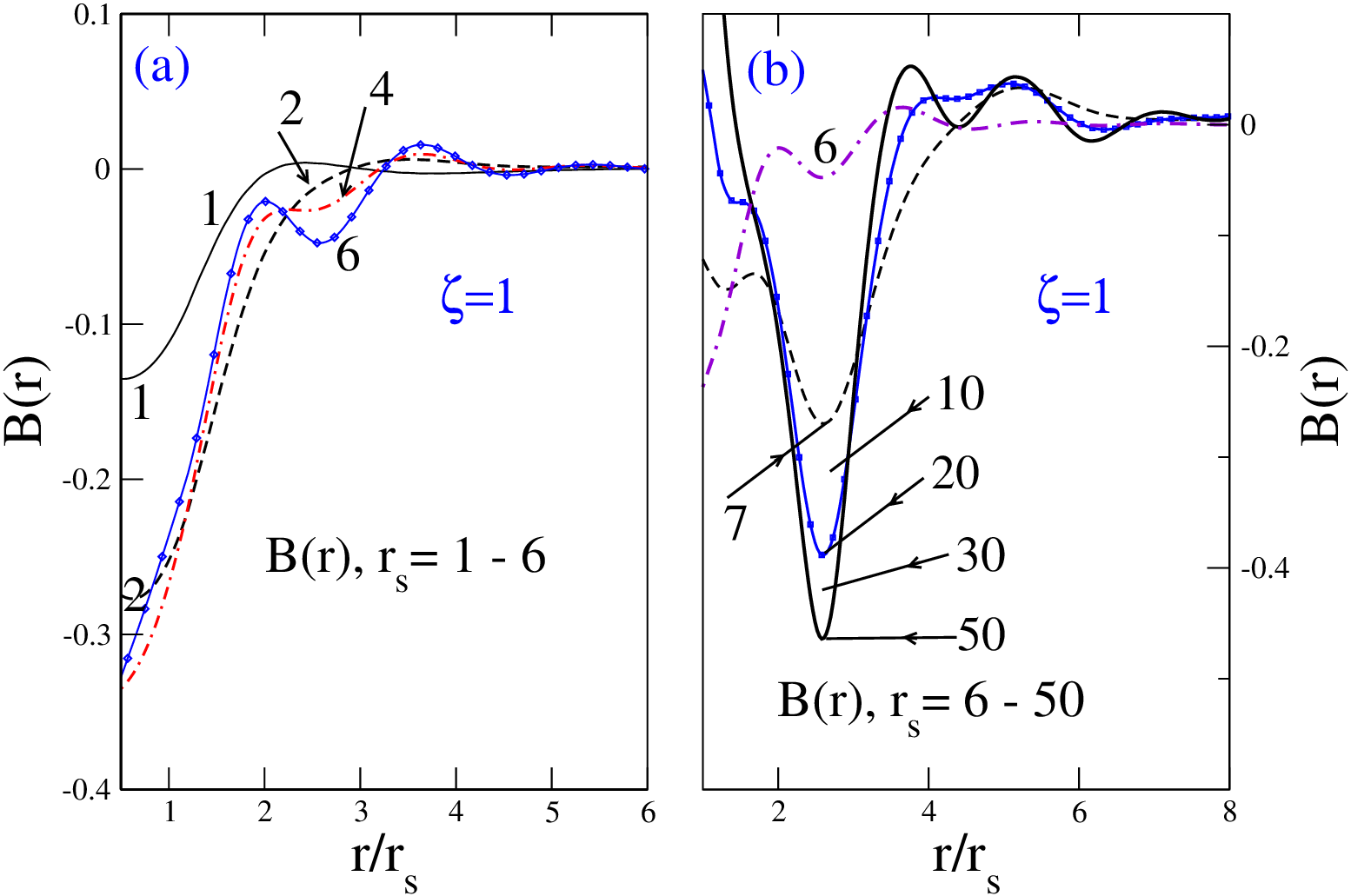}
 \caption
{(Online color)The 2D Coulomb bridge functions needed for
 a classical description of the
pair-distribution function of a fully spin-polarized ($\zeta=1$)
2D electron fluid, for $r_s=1$ to 50. Panel (a) shows
the weak to moderately coupled regime. Panel (b) shows that
the nature of the $B(r)$
changes rapidly near $r_s=6$ and $r_s=7$, when the second
 trough in the
PDF begins to develop. 
The PDFs are insensitive to the behaviour of these
 functions for $r/r_s <\sim 1.3$.
}
\label{bridgefig-z1}
\end{figure}

The extraction procedure for the Coulomb bridge functions $B_{i,j}(r)$
from the QMC-PDFs, i.e., $g_{ij}(r)$ is found to be very efficient, and the iterative
inclusion of the extracted bridge functions leads to rapid 
convergence in reproducing the target $g_{ij}(r)$. 
 It is easiest to consider
a fully spin-polarized (i.e., $\zeta=1$) electron fluid as it is a one-
component system, with just one bridge function. A set of bridge functions
for a range of $r_s$ values which reproduce the QMC pair-distribution
 functions at $\zeta=1$
when used in CHNC are shown in Fig.~\ref{bridgefig-z1}.
It is found that as $r_s$ increases from 0.3 (not shown) to unity,
the $B(r)$ remains more or less unchanged. From then onwards,
esp. after $r_s=3$, the development of the first peak in the PDF
produces an oscillatory structure in $B(r)$. This trend continues
 till about $r_s=6$. However, now the deepening of the second trough in the PDF
leads to a complete qualitative change in the the form of the bridge function,
as seen from the panel (b) of Fig.~\ref{bridgefig-z1}. The new
deep trough seen near
$x=r/r_s \sim 3$ is already there as a weak trough in the moderately coupled
fluids of panel (a). In panel (a), the deep trough
is in the $r/r_s <1$ region and it is not of much importance
in the HNC equation. In panel (b), when the deep trough near $x\sim3$ develops,
the small $x$ region rises rather steeply. But this rise has only a small
numerical significance in the HNC, since the pair potential and
the Pauli potential 
become very large and dominant for small $x$. In contrast, the structure
 in $B(r)$
for larger $x$ becomes significant,  as the pair-potential effects fall off with increasing $x$.
These results show that
the short-range structure of the fluid undergoes significant and subtle changes
in the regime $r_s>6$. Previous studies of the local-field corrections to the
response function of the 2D electron liquid had unraveled
interesting characteristics~\cite{lfc2d, Atwal2003} which may well
be related to the onset of strongly correlated clusters for $r_s>6$.

Another system which effectively reduces to a one-component
 system is the paramagnetic
 electron fluid ($\zeta=0$) with equal amounts of up-spin, and
down-spin species. As a test of spin insensitivity,
we can use the bridge functions determined
 from the $\zeta=1$ case
for the $\zeta=0$ case and see if the CHNC-PDFs reproduce the QMC-PDFs. This is in fact very nearly the case, showing that the spin-polarization dependence of $B(r)$ is
very small. This also shows that, if desired, one may introduce just
one bridge function $B(r,\zeta)$ for all $i,j$ components, where the latter is based
on linearly interpolating between $B(r,\zeta=1)$ and $B(r,\zeta=0)$, since they are
very similar. The differences are found 
 in the near $r\sim r_s$ region
close to the first peak, where the Pauli-exclusion effects, diffraction effects
etc., are comparable to the Coulomb repulsion effects (see the discussion of
 Fig.~\ref{att-drum-chnc.fig}).   
%
%
\section{Discussion}
\label{discus-sec}

Currently, the most sophisticated QMC calculations for the quantum 2D electron liquid
are those of Drummond et al. In practice, the differences between those of
Drummond et al., and those of Attaccalite et al., are irrelevant except for 
 dealing with very small energy differences
 indicative of a phase transition from the paramagnetic to the ferromagnetic state. The presence or absence
 of a ferromagnetic transition would also modify the predictions
regarding the enhancement of $g^*$.
The lack of such a
transition in the results of Drummond et al., further illustrate the difficulties inherent in assuming that a given set of QMC data are essentially the definitive result. Given the very slight energy differences between the two phases, any technical improvements can modify the conclusions. Since CHNC is based on using some of the QMC data as inputs, and then applying CHNC to other instances, the conclusions of CHNC also get modified as the input QMC data are modified. Thus, in our study of the
two-valley 2D electron gas~\cite{2v2d}
 found in Si metal-oxide field-effect transistors (MOSFETS), 
we used hard-sphere bridge functions, and the classical-fluid temperatures
based on the QMC correlation energies of Tanatar and Ceperley
as inputs to the CHNC. It was found that the two-valley systems becomes critical at sufficiently high $r_s$ values, when the effective mass and the spin susceptibility become very large. This property is no longer found to be the case when more accurate bridge functions are used.
\begin{figure}[!h]
\includegraphics*[width=8.0 cm, height=6.4 cm]{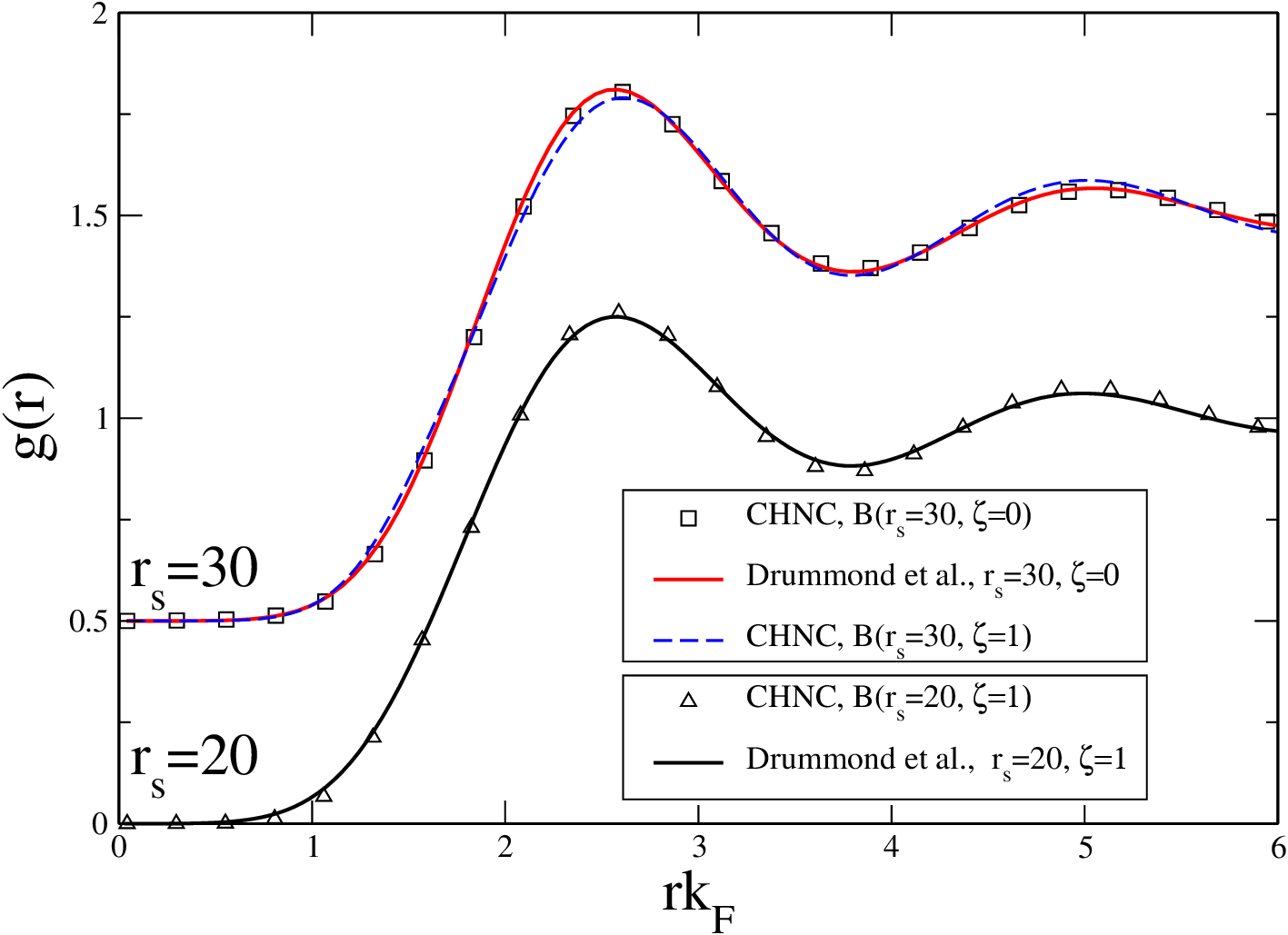}
 \caption
{(Online color) A comparison of the 2D electron-liquid
PDFs of Drummond et al~\cite{Drum2009} with those obtained
from the CHNC, using the bridge functions iterated from
Eq.~\ref{iter-bridge}. In the case of $r_s=30$, the Drummond et al.
data are for $\zeta=0$, and the effect of using the $\zeta=1$
bridge function on the PDF (dashes) in CHNC instead of the
 correct $B(r)$ at $\zeta=0$  is also
shown (PDF,Boxes). The PDFs at $r_s=30$ have been displaced
 upwards for clarity.
}
\label{att-drum-chnc.fig}
\end{figure}
 
The pair distribution functions calculated from CHNC using the bridge functions extracted from the Gori-Giorgi et al. fit to the Attaccalite $g(r)$ are, of course, essentially identical to the target $g(r)$.
Thus a comparison of CHNC-$g(r)$ using the fitted $B(r)$
is a check on our fitting process as well as an indirect
 comparison of Refs.~\cite{atta} and \cite{Drum2009}. It also enables us to check on the applicability of the $B(r)$ obtained from the
polarized-2D system for predictions on the unpolarized system.
In fig.~\ref{att-drum-chnc.fig} we
compare CHNC-$g(r)$ with numerical data from 
Drummond et al.~\cite{DrumAck}, for $r_s=20, \zeta=1$, and $r_s=30, \zeta=0$. In the latter case, we have also calculated the CHNC PDFs using the appropriate bridge function
 $B(r,\zeta=0)$, and also the inappropriate $B(r,\zeta=1)$ to display the
relative insensitivity of the PDF to any spin-polarization dependence in
$B(r)$. 

It should also be noted that we have not modified the quantum temperature
$T_{q}$ of the classical map given in Ref.~\cite{prl2} in extracting the bridge 
functions. In the classical map of Balutay et al., a pure-HNC procedure and a different $T_q$  were used. If it were extended to include the bridge terms, qualitatively very similar results are obtained for the bridge functions which are now slightly different
functions.

 Once a family of bridge functions for each $r_s$ and $T_q$ is obtained, the $g(r)$ for each $r_s$ as well as all the smaller $r_s$
data are used in the adiabatic-connection formula for the correlation energy $E_c$. Then the QMC
value of $E_c$ will be accurately recovered, and hence we need not revise the preset quantum temperature $T_q$ that was used in the
classical map. That is, given a $T_q$ of a classical map of the CHNC type, the use of the Coulomb bridge function based on that $T_q$
instead of the original bridge functions used when forming the $T_q$, does not require any refitting
the $T_q$. This is a very convenient conservation property for
implementing CHNC calculations with improved bridge functions.   

The advantage of the CHNC procedure over those of QMC is its simplicity of implementation, as well as its easy applicability to finite-$T$, finite-$\zeta$ situations. Thus CHNC studies of multi-valley systems (e.g., Si-MOSFETs), nanostructures etc., can be attempted. 
Finite-$T$ studies become possible if the $B(r)$ remain valid
 even for finite-$T$, at least in the near $T=0$ region. An
estimate of the transferability of the $T=0$ bridge functions to
finite-$T$ may be obtained by noting that temperature acts to
level out the oscillations in the PDFs, while the $B(r)$ attempts
to sharpen the oscillations.

 Some years ago we showed~\cite{prl3} that very interesting
spin-dependent effects arise in the $T=0$ to $T=E_F$ region due to
 the interplay of
spin, onset of partial degeneracy, Coulomb correlations and temperature effects. Those studies were carried out with the hard-sphere bridge functions available at that time. Such temperature-dependent studies are still beyond the reach of QMC methods.
The accuracy of those earlier $B(r)_{HS}$-based
studies at finite-$T$ can now be further examined using the new bridge functions.

Finite temperature studies of the free energy of the 2D gas as a function of density and spin polarization can also be used to determine Fermi-liquid parameters like the effective mass~\cite{Holzmann,drum2} and the spin susceptibility. In CHNC we directly
determine $m^*$ from the second density derivative of the
excess free energy with respect to the density~\cite{gstrmstr}.
Thus, the availability of these new Coulomb bridge functions which accurately describe the $T=0$ PDFs provide greater confidence in the results that may be obtained for the Fermi-liquid parameters in future studies.     

In conclusion, we have presented an {\it accurate classical representation} of the pair-distribution functions of the quantum 2D electron fluid at $T=0$, for arbitrarily strong coupling and arbitrary spin polarizations.

\end{document}